\documentclass[runningheads]{llncs}
\usepackage{graphicx}
\usepackage{amsmath}
\usepackage{amssymb}
\usepackage{caption}
\usepackage{multirow}
\usepackage{booktabs}
\usepackage{makecell}
\usepackage{hyperref}
\usepackage{booktabs, tabularx}
\newcommand{\beginsupplement}{%
        \setcounter{table}{0}
        \renewcommand{\thetable}{S\arabic{table}}%
        \setcounter{figure}{0}
        \renewcommand{\thefigure}{S\arabic{figure}}%
     }

\begin{document}
\title{Biomechanics-informed Neural Networks for Myocardial Motion Tracking in MRI}
\titlerunning{Biomechanics-informed Neural Networks}

\author{Chen Qin\inst{1,2}, Shuo Wang\inst{3}, Chen Chen\inst{1}, Huaqi Qiu\inst{1}, Wenjia Bai\inst{3,4}, and Daniel Rueckert\inst{1}}
\institute{Department of Computing, Imperial College London, London, UK \email{c.qin15@imperial.ac.uk}\\
\and Institute for Digital Communications, School of Engineering, University of Edinburgh, Edinburgh, UK\\
\and
Data Science Institute, Imperial College London, London, UK\\
\and 
Department of Brain Sciences, Imperial College London, London, UK}
\authorrunning{C. Qin et al.}

\maketitle              
\begin{abstract}
Image registration is an ill-posed inverse problem which often requires regularisation on the solution space. In contrast to most of the current approaches which impose explicit regularisation terms such as smoothness, in this paper we propose a novel method that can implicitly learn biomechanics-informed regularisation. Such an approach can incorporate application-specific prior knowledge into deep learning based registration. Particularly, the proposed biomechanics-informed regularisation leverages a variational autoencoder (VAE) to learn a manifold for biomechanically plausible deformations and to implicitly capture their underlying properties via reconstructing biomechanical simulations. The learnt VAE regulariser then can be coupled with any deep learning based registration network to regularise the solution space to be biomechanically plausible. The proposed method is validated in the context of myocardial motion tracking on 2D stacks of cardiac MRI data from two different datasets. The results show that it can achieve better performance against other competing methods in terms of motion tracking accuracy and has the ability to learn biomechanical properties such as incompressibility and strains.
The method has also been shown to have better generalisability to unseen domains compared with commonly used L2 regularisation schemes. 

\end{abstract}
\section{Introduction}
Medical image registration plays a crucial role in inferring spatial transformation of anatomical structures, and has been successfully used for various applications such as multi-modal image fusion, detection of longitudinal structural changes and analysis of motion patterns. Due to the intrinsic ill-posedness of the registration problem, there exist many possible solutions, i.e. spatial transformations, to register between images. To ensure deformation to be unique and physiological plausible, regularisation techniques are often employed to convert the ill-posed problem to a well-posed one by adding a regularisation term with desired properties to the registration objective function. Specifically, image registration between a pair of images $I_s$, $I_t$ can be formulated as a minimisation problem:
\begin{equation}
\underset{\Phi \in \mathcal{D}(\Omega)}{\mathrm{argmin}}\  \mathcal{L}_{sim}(I_t, I_s \circ \Phi) + \mathcal{R}(\Phi)
\end{equation}
where $\mathcal{L}_{sim}$ stands for the image dissimilarity measure, $\Omega$ is the image domain, $\Phi$ denotes the transformation from source image $I_s$ to target image $I_t$, $\mathcal{D}(\Omega)$ is the group of feasible transformations and $\mathcal{R}$ is a regularisation term. Conventionally, the regularisation is imposed explicitly on $\mathcal{R}$ or $\mathcal{D}(\Omega)$ with assumptions on the deformation field such as smoothness, diffeomorphism or incompressibility. Though these assumptions are generic, there is often a lack of incorporation of application-specific prior knowledge to inform the optimisation.

In this work, instead of imposing an explicit regularisation, we propose a biomechanics-informed neural network for image registration which can implicitly learn the behaviour of the regularisation function and incorporate prior knowledge about biomechanics into deep learning based registration. Specifically, the learning-based regularisation is formulated using a variational autoencoder (VAE) that aims to learn a manifold for biomechanical plausible deformations, and this is achieved via reconstructing biomechanically simulated deformations. The learnt regulariser is then coupled with a deep learning based image registration, which enables the parameterised registration function to be regularised by the application-specific prior knowledge, so as to produce biomechanically plausible deformations. The method is evaluated in the context of myocardial motion tracking in 2D stacks of cardiac MRI data. We show that the proposed method can achieve better performance compared to other competing approaches. It also indicates great potential in uncovering properties that are represented by biomechanics, with a more realistic range of clinical parameters such as radial and circumferential strains, as well as better generalisation ability.

\paragraph{Related Work}
Deformable image registration has been widely studied in the field of medical image analysis. The majority of the methods explicitly assume the smoothness on the deformation, such as introducing a smoothness penalty in the form of L2 norm or bending energy to penalise the first- or second-order derivatives of the deformation \cite{sotiras2013deformable}, or parameterising the transformation via spline-based models in a relatively low dimensional representation space \cite{rueckert1999nonrigid}. To take into account the invertibility of the transformation, many diffeomorphic registration methods have been proposed to parameterise the displacement field as the integral of a time-varying velocity field or stationary velocity field (SVF) \cite{beg05,ashburner07}, which mathematically guarantees diffeomorphism. Volumetric preservation is also a characteristic that is often desired for soft tissue tracking such as myocardium tracking to ensure that the total volume of myocardium keeps constant during image registration \cite{fidon2019incompressible,mansi2011ilogdemons,shi2012comprehensive}. Such incompressible registration methods either relax $\text{det}(J_{\Phi}(x))=1$ as a soft constraint \cite{shi2012comprehensive} (where $J_{\Phi}$ is the Jacobian matrix of the transformation $\Phi$), or use specific parameterisation for the displacement field or SVF \cite{fidon2019incompressible,mansi2011ilogdemons}. In addition, population statistics have also been investigated to incorporate prior knowledge for the registration process \cite{khallaghi2015statistical,rueckert2003automatic}. 

More recently, deep learning based image registration has been shown to significantly improve the computational efficiency. Similarly, most deep learning approaches regularise deformation fields with a smoothness penalty \cite{fan2018adversarial,qin2018joint,qin2018undersampled}, or parameterise transformations with time-varying or stationary velocity fields \cite{balakrishnan2019voxelmorph,bone2019learning,krebs2019learning,qin2019unsupervised}. Besides, population-based regularisation has been designed to inform registration with population-level statistics of the transformations \cite{bhalodia2019cooperative}. Adversarial deformation regularisation was also proposed which penalised the divergence between the predicted deformation and the finite-element simulated data \cite{hu2018adversarial}. In contrast to the existing deep learning based registration work, our method proposes to implicitly learn the regularisation function via a biomechanics-informed VAE. This aims to capture deformation properties that are biomechanically plausible, with a particular focus on the application of myocardial motion tracking.

\section{Method}
\begin{figure*}[!t]
\centering
\includegraphics[width=.85\linewidth]{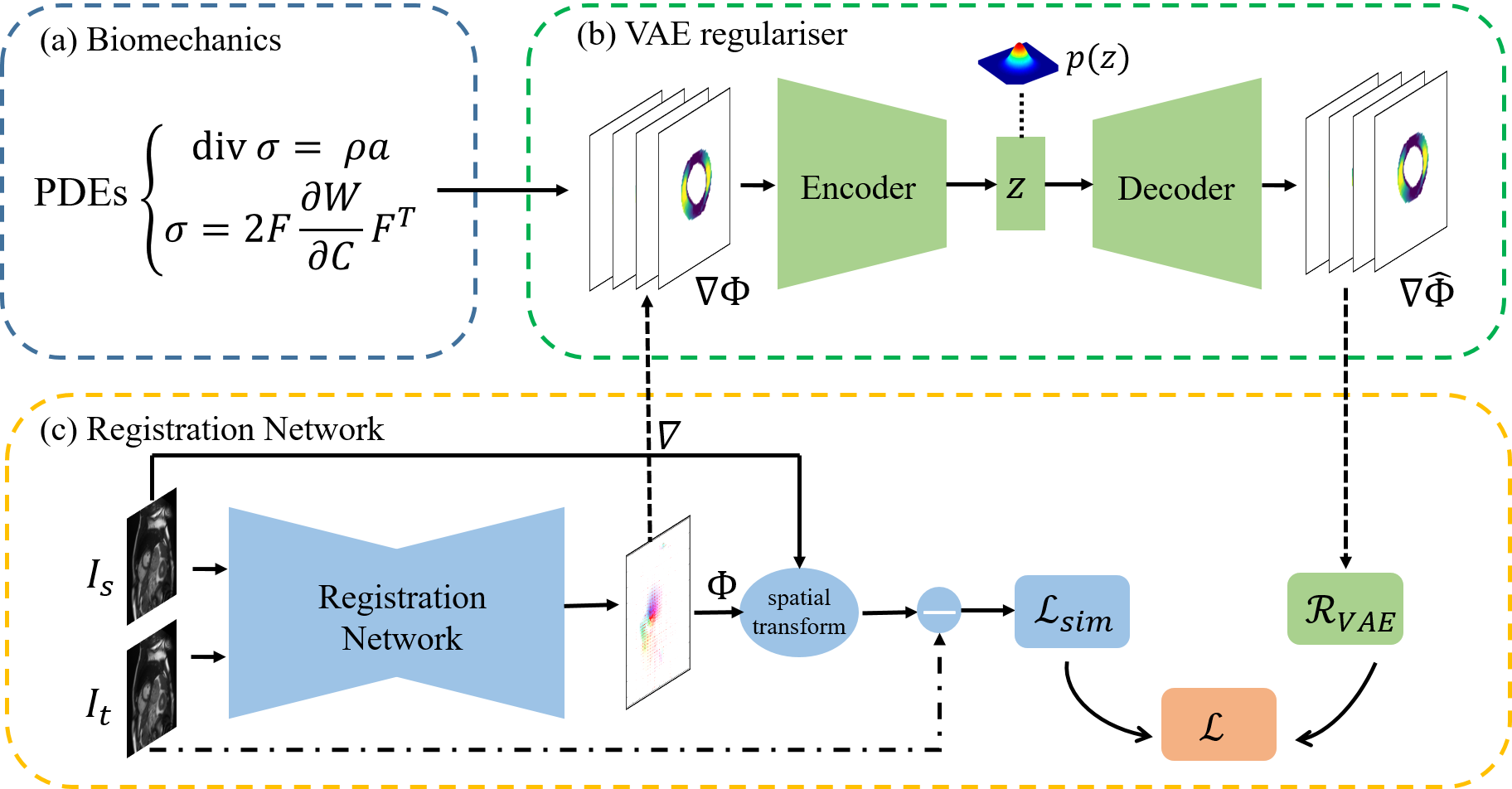}
\caption{Illustration of biomechanics-informed neural network for image registration: (a) Biomechanical simulations are generated according to partial differential equations (PDEs). (b) A VAE regulariser is learnt via reconstructing the first-order gradients of simulated displacement fields. (c) The registration network is regularised by biomechanics via incorporating the VAE regularisation term.}
\label{fig:framework}
\end{figure*}

The proposed biomechanics-regularised registration network is illustrated in Fig.\ref{fig:framework}, which mainly consists of three components. First, biomechanical simulations of deformation are generated according to equations which guarantee the physical properties of the deformation (Sec.~\ref{sec:biomechanical simulations}). Second, VAE is leveraged to learn the probability distribution of the simulated deformations, which implicitly captures the underlying biomechanical properties (Sec.~\ref{sec:VAE_reg}). Finally, the learnt VAE then acts as a regularisation function for the registration network, which regularises the solution space and helps to predict biomechanically plausible deformation without the need for any explicit penalty term (Sec.~\ref{sec:registration network}).

\subsection{Biomechanical Simulations}
\label{sec:biomechanical simulations}
To enable a learning-based biomechanically informed regulariser, we first propose to generate a set of deformation fields using a biomechanical simulation. From the perspective of solid mechanics, the modelling of myocardial motion can be considered as a plane-strain problem with the following governing equation \cite{hunter1988analysis}:
\begin{equation}
{\rm div}\ \sigma = \rho a 
\end{equation}
Here $\sigma$ is the stress tensor, ${\rm div}\ \sigma$ denotes the divergence of the stress tensor, $\rho$ is the material density and $a$ is the acceleration. This describes the equilibrium within an elastic body. To solve the deformation, the constitutive law is provided, assuming a hyper-elastic material behaviour:
\begin{equation}
\sigma = 2F  \frac{\partial W}{\partial C} F^T
\end{equation}
where $F$ is the deformation gradient, $C$ is the right Cauchy–Green deformation tensor and $W$ stands for the strain energy density function (SEDF). We adopt a neo-Hookean constitutive model with the shear modulus of 36.75 kPa from experiments ~\cite{zhu2014numerical}, which guarantees the material incompressibility. The above equations can be solved using the finite element method (FEM) with appropriate boundary conditions. For a given myocardial segmentation at the end-systolic (ES) frame, we simulate the artificial 2D deformation in a one cardiac cycle by adjusting the blood pressure (Supple. Fig. \ref{fig:S1}).

\subsection{VAE regularisation}
\label{sec:VAE_reg}
Instead of explicitly specifying a regularisation term that is suitable for the ill-posed registration problem, here we propose to implicitly learn the prior knowledge from biomechanics as regularisation. In particular, the VAE formulation \cite{kingma2013auto} is leveraged here as a learning-based regulariser to model the probability distribution of biomechanically plausible deformations. The VAE encodes an input and returns a distribution over the latent space, which enables the inverse problem to move through the latent space continuously to determine a solution. 

In detail, the VAE regulariser is trained to reconstruct the first-order derivative of the biomechanically simulated deformations to remove any effects of rigid translation. Let us denote the deformation in 2D space by $\Phi =\left[u, v\right] \in \mathrm{R}^{2 \times M \times N}$, where $u$ and $v$ denote the displacements along $x$- and $y$- directions respectively. The first-order gradients of the deformation field can be represented as $\triangledown \Phi = \left[\frac{\delta u}{\delta x}, \frac{\delta u}{\delta y}, \frac{\delta v}{\delta x}, \frac{\delta v}{\delta y}\right] \in \mathrm{R}^{4 \times M \times N}$, where $M$ and $N$ denote the spatial dimensions. The reconstruction loss of the VAE is formulated as, 
\begin{equation}
    \mathcal{R}_\mathrm{VAE}(\triangledown \Phi) = \|\triangledown \Phi - \triangledown \hat{\Phi}\|_{2}^{2} + \beta \cdot D_{KL}(q_{\theta}(\mathbf{z}|\triangledown {\Phi})\|p(\mathbf{z})),
\end{equation}
where $\triangledown \hat{\Phi}$ denotes the reconstruction from input $\triangledown \Phi$, $\mathbf{z}$ denotes the latent vector encoded by VAE, $p(\mathbf{z}) \sim \mathcal{N}(0,I)$ denotes the prior Gaussian distribution, $q_{\theta}$ denotes the encoder parameterised by $\theta$ and $D_{KL}$ represents the Kullback-Leibler divergence. $\beta$ is a hyperparameter that controls the trade-off between the reconstruction quality and the extent of latent space regularity.

\subsection{Biomechanics-informed registration network}
\label{sec:registration network}
As shown in Fig.\ref{fig:framework}(c), the registration network aims to learn a parameterised registration function to estimate dense deformation fields between a source image $I_s$ and a target image $I_t$. As the ground truth dense correspondences between images are not available, the model thus learns to track the spatial features through time, relying on the temporal intensity changes as self-supervision. The image dissimilarity measure thus can be defined as $\mathcal{L}_{sim}(I_t, I_s, \Phi)=\|I_t-I_s \circ \Phi\|^2_2$, which is to minimise the pixel-wise mean squared error between the target image and the transformed source image. 

To inform the registration process with prior knowledge of biomechanics, we propose to integrate the VAE regularisation into the registration network. This aims to regularise the viable solutions to fall on the manifold that is represented by the latent space of VAE, which thereby enables the network to predict biomechanically plausible deformations. This is advantageous over the explicit regularisation terms (e.g. smoothness penalty), as the data-driven regulariser implicitly captures realistic and complex properties from observational data that cannot be explicitly specified. In addition, the VAE loss $\mathcal{R}_{\mathrm{VAE}}$ provides a quantitative metric for determining how biomechanically plausible the deformation is. Solutions close to the learnt VAE latent manifold will produce a low $\mathcal{R}_{\mathrm{VAE}}$ score, whereas solutions far from the manifold will give a higher score. The final objective function for training the registration network is then formulated as:
\begin{equation}
\label{loss}
    \mathcal{L}=\mathcal{L}_{sim}(I_t, I_s \circ \Phi) + \alpha \cdot \mathcal{R}_{\mathrm{VAE}}(\triangledown \Phi \odot \mathbf{M}).
\end{equation}
Here $\mathbf{M}$ is a binary myocardial segmentation mask which is provided to only regularise the region of interest, and $\odot$ represents the Hadamard product. A hyperparameter $\alpha$ is introduced here, which trades-off image similarity and the physical plausibility of the deformation.

\section{Experiments and Results}
\begin{figure}[!h]
\centering
\includegraphics[width=\linewidth]{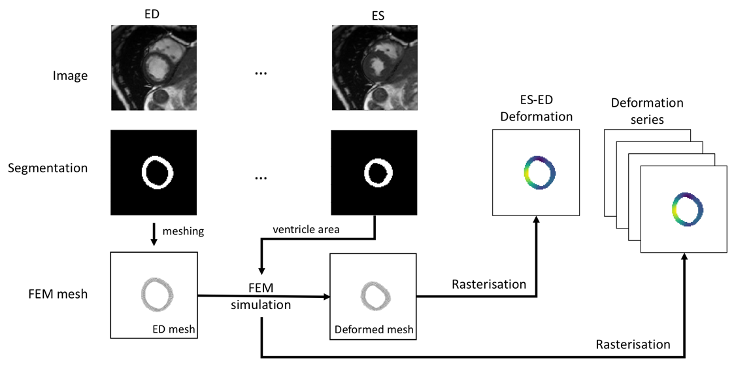}
\caption{Biomechanical simulation of myocardium deformation. The segmentation at ED frame is transformed into a quadrilateral mesh for finite element method (FEM) simulation. An artificial pressure was assumed to be applied on the inner border. The ES frame is assumed as the undeformed state and the ED frame is deformed under pressure $P_{ED}$. The $P_{ED}$ value is determined through an inverse FEM scheme matching the ES ventricle area. The artificial deformation field is rasterised as an artificial ES-ED deformation sample. By uniform sampling $P_{ED}$ in [0, $P_{ED}$] for 50 times, a series of artificial deformation field in one cardiac cycle is obtained.}
\label{fig:S1}
\bigskip
\bigskip
\centering
\includegraphics[width=.95\linewidth]{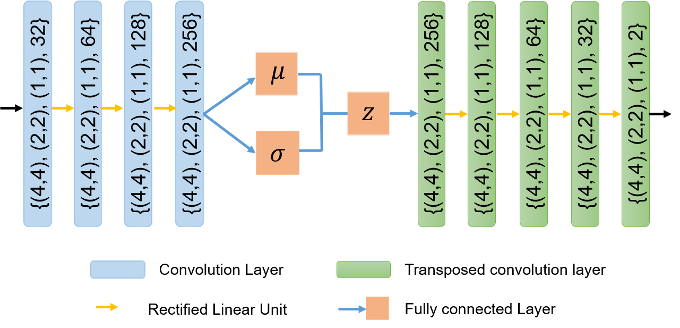}
\caption{Detailed network architecture of VAE. Numbers inside CNN layer are {kernel size, stride, padding, number of filters}.}
\label{fig:S2}
\end{figure}

Experiments were performed on 300 short-axis cardiac cine magnetic resonance image (MRI) sequences from the publicly available UK Biobank (UKBB) dataset \cite{petersen2017reference}. Each scan consists of 50 frames and each frame forms a 2D stack of typically 10 image slices. Segmentation masks for the myocardium were obtained via an automated tool provided in \cite{bai2018automated}. In experiments, we randomly split the dataset into training/validation/testing with 100/50/150 subjects respectively. Image intensity was normalised between 0 and 1. Data used for biomechanical simulations were from a separate data set in UKBB consisting of 200 subjects. The detailed VAE architecture is described in the Supple. Fig. \ref{fig:S2} (the latent vector dimension was set to 32, and $\beta = 0.0001$). The network architecture for registration was adopted from \cite{qin2018joint} due to its proven effectiveness in motion tracking. 
A learning rate of 0.0001 with Adam optimiser was employed to optimise both the VAE and the registration network.

\begin{figure*}[!t]
\centering
\includegraphics[width=\linewidth]{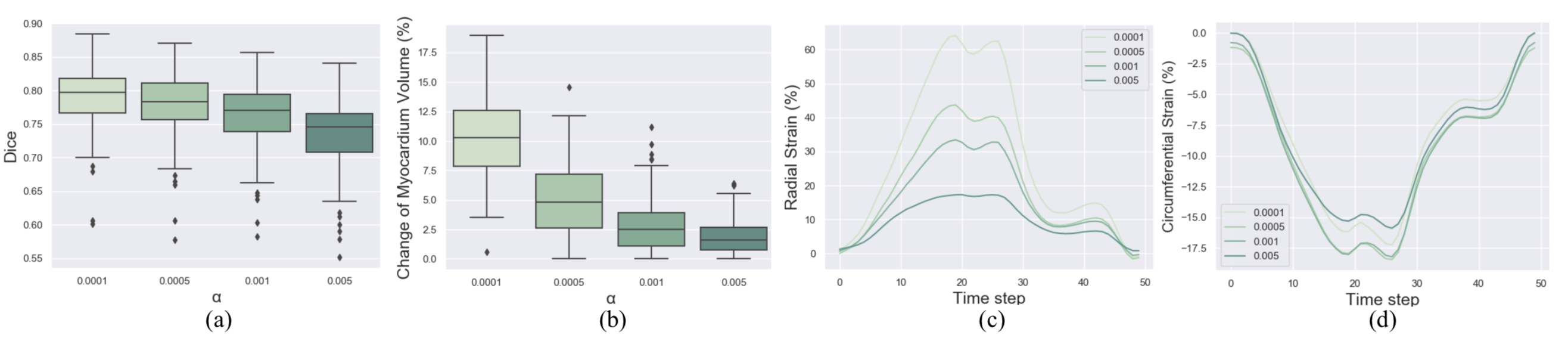}
\caption{Study on effects of VAE regularisation. By varying $\alpha$, we show the corresponding effects on (a) Dice coefficient and (b) Change of myocardial volume between ES and ED frames; (c) Radial strain and (d) Circumferential strain across the cardiac cycle.}
\label{fig:ablation}
\end{figure*}

\paragraph{Effect of VAE regularisation.} An ablation study to investigate the effect of $\alpha$ (the weight for VAE regularisation) was performed. Here $\alpha$ is set to values of \{0.0001, 0.0005, 0.001, 0.005\}, and the performance was evaluated using three quantitative measures, i.e. Dice overlap metric, change of myocardial volume and myocardial strain, as shown in Fig.\ref{fig:ablation}. In detail, for results in Fig.\ref{fig:ablation}(a)(b), motion was estimated between ES and end-diastolic (ED) frames of the cardiac image sequence, where ES frame is warped to ED frame with the estimated transformation. It can be observed that as $\alpha$ increases, the change for myocardial volume is reduced which means better volume preservation, and the observed radial and circumferential strains (Fig.\ref{fig:ablation}(c)(d)) are more plausible and within the range that has been reported in the literature \cite{ferdian2020fully}. However, this is at the expense of the decrease of registration accuracy in terms of the Dice metric. In addition, since the motion estimation was performed in stack of 2D slices, the incompressibility of the myocardium can not be guaranteed due to the out-of-plane motion. Nevertheless, results here indicate that the proposed method has the potential and ability in implicitly capturing such biomechanical properties.

\paragraph{Comparison Study.}
The proposed registration network was then compared with other state-of-the-art motion estimation methods, including free-form deformation (FFD) with volumetric preservation (VP) \cite{rohlfing2001intensity}, diffeomorphic Demons (dDemons) \cite{vercauteren2007non}, and a deep learning based registration which uses the L2 norm to penalise displacement gradients (DL+L2) \cite{qin2018joint}. A quantitative comparison results are shown in Table \ref{motion_evaluation}, where the performance was evaluated in terms of the mean contour distance (MCD), Dice, and the mean absolute difference between Jacobian determinant $|J|$ and 1 over the whole myocardium, denoted as $||J|-1|$, to measure the level of volume preservation. In experiments, we separate these evaluations on different representative slices, i.e., apical, mid-ventricle, and basal slices. For fair comparison, hyperparameters in all these methods under comparison are selected to ensure that they can achieve an average of $||J|-1|$ to be in a range around 0.1 or slightly higher, and $\alpha$ in the proposed method is thus chosen as 0.001. From Table \ref{motion_evaluation}, it can be observed that the proposed method can achieve better or comparable performance on all these slices in terms of both registration accuracy (MCD and Dice) and volume preservation ($||J|-1|$) compared with other methods. The performance gain is especially significant on more challenging slices such as basal slices (MCD: 2.280 (proposed) vs 2.912 (DL+L2); Dice: 0.725 vs 0.619), where prior knowledge of physically plausible deformation can have more impact on informing the alignment.

\begin{table*}[!t]
  \centering
  \caption{Comparisons of motion estimation performance between FFD with volume preservation (FFD+VP), diffeomorphic Demons (dDemons), deep learning based registration with L2 norm regularisation (DL+L2), and the proposed method with $\alpha=0.001$. Results are reported as mean (standard deviation). Lower MCD and higher Dice indicates better accuracy. Lower $||J|-1|$ indicates better volume preservation.}
  \label{motion_evaluation}
\scalebox{0.9}
 { \begin{tabular}{cccccccccc}
  
    \toprule
\multirow{2}*{Method} & \multicolumn{3}{c}{Apical} & \multicolumn{3}{c}{Mid-ventricle}& \multicolumn{3}{c}{Basal}\\
\cline{2-10}
 & {MCD} & {Dice}   & $||J|-1|$  & {MCD} & {Dice}   & $||J|-1|$   & {MCD} & {Dice}   & $||J|-1|$  \\
  \midrule
  {FFD+VP \cite{rohlfing2001intensity}} & \makecell{2.766\\(1.240)} &\makecell{0.598\\(0.186)} &\makecell{0.207\\(0.045)} & \makecell{2.220\\(0.750)}&\makecell{0.725\\(0.077)}&\makecell{0.186\\(0.033)} &\makecell{3.958\\(1.925)} &\makecell{0.550\\(0.199)} &\makecell{0.176\\(0.028)}\\
  \midrule
  
 {dDemons \cite{vercauteren2007non}} & \makecell{2.178\\(1.079)} & \makecell{0.651\\(0.186)} & \makecell{0.133\\(0.044)} & \makecell{1.603\\(0.629)} & \makecell{0.784\\(0.072)} & \makecell{0.148\\(0.036)} & \makecell{3.568\\(1.482)} & \makecell{0.560\\(0.176)} & \makecell{0.132\\(0.029)}\\
 \midrule
DL+L2 \cite{qin2018joint} & \makecell{2.177\\(1.119)}&\makecell{0.639\\(0.196)}&\makecell{\textbf{0.087}\\(0.028)}&\makecell{{1.442}\\(0.625)}&\makecell{\textbf{0.806}\\(0.063)}&\makecell{0.104\\(0.042)}&\makecell{2.912\\(1.384)}&\makecell{0.619\\(0.178)}&\makecell{0.149\\0.047} \\
\midrule
\midrule
\makecell{Proposed \\ $\alpha=0.001$} & \makecell{\textbf{1.734}\\(0.808)}&\makecell{\textbf{0.684}\\(0.160)} &\makecell{0.126\\(0.071)}&\makecell{\textbf{1.417}\\(0.402)} & \makecell{0.796\\(0.059)} &\makecell{\textbf{0.081}\\(0.020)} & \makecell{\textbf{2.280}\\(1.274)} & \makecell{\textbf{0.725}\\(0.149)} &\makecell{\textbf{0.124}\\(0.037)}\\
    \bottomrule
  \end{tabular}}
\end{table*}

In addition, we further evaluated the performance of the proposed method on myocardial strain estimation, as shown in Table \ref{strain_evaluation}. The peak strain was calculated between ES and ED frames on cine MRI using the Lagrangian strain tensor formula \cite{elen2008three}, and was evaluated on apical, mid-ventricle and basal slices separately. To better understand the strains, we compared the predictions from cine MRI with reference values obtained from tagged MRI \cite{ferdian2020fully}, which is regarded as reference modality for cardiac strain estimation. The reference values were taken from Table 2 in \cite{ferdian2020fully} whose results were derived also from UKBB population (but on a different subset of subjects).
Compared with other methods, the proposed approach can achieve radial and circumferential strain in a more reasonable value range as defined by the reference. Particularly, basal slice strain and circumferential strain are very consistent with the reference values reported. Note that evaluation of radial strain is challenging, and there is a poor agreement in it even for commercial software packages \cite{cao2018comparison}. In this case, incorporating biomechanical prior knowledge may benefit motion tracking and lead to more reasonable results.
Besides, the Jacobian determinant of deformation across the entire cardiac cycle on one subject is presented in Fig. \ref{fig:jacobian}. Here the proposed biomechanics-informed method is compared with the one using L2 norm, and it shows better preservation of the tissue volume at different cardiac phases, with Jacobian determinant being close to 1. 
This implies that the proposed learning based regularisation informed by biomechanics is able to capture and represent biomechanically realistic motions. A dynamic video showing the qualitative visualisations is also presented in supplementary material.

\begin{table*}[!t]
  \centering
  \caption{Comparisons of peak strain values (\%) obtained from different methods at different slices. Reference values are derived from tagged MRI based on a similar UK Biobank cohort \cite{ferdian2020fully}. RR: peak radial strain; CC: peak circumferential strain. Bold results indicate results that are closest to the reference values.}
  \label{strain_evaluation}
\setlength{\tabcolsep}{3pt}
 { \begin{tabular}{ccccccccccc}
  
    \toprule
\multirow{2}*{Method} & \multicolumn{2}{c}{FFD+VP\cite{rohlfing2001intensity}} & \multicolumn{2}{c}{dDemons\cite{vercauteren2007non}}& \multicolumn{2}{c}{DL+L2\cite{qin2018joint}}&\multicolumn{2}{c}{Proposed} & \multicolumn{2}{c}{Reference\cite{ferdian2020fully}}\\
\cline{2-11}
 & {RR} & {CC}   & RR  & CC & RR   & CC   & RR & CC   & RR & CC  \\
  \midrule
  {Apical} & 14.9 & -7.4 & 30.0 & -14.2 & \textbf{26.5} & -14.1 & 32.5 & \textbf{-19.7} & \textit{18.7} & \textit{-20.8} \\
  \midrule
 {Mid-ventricle} & 12.8 & -7.82 & 37.6 & -12.7 & 35.9 & -14.0 & \textbf{33.1} & \textbf{-18.1} & \textit{23.1} & \textit{-19.6}\\
 \midrule
Basal & 12.4 & -5.9 & 22.8 & -10.9 & 31.6 & -13.0 & \textbf{24.5} & \textbf{-16.4} & \textit{23.8} & \textit{-16.7}\\
    \bottomrule
  \end{tabular}}
\end{table*}

\paragraph{Generalisation Study.}
We further performed a generalisation study of the proposed method against the L2 norm regularised network. Specifically, we deployed models trained on the UKBB dataset and directly tested them on 100 ACDC data \cite{bernard2018deep}. The tested results are shown in Table \ref{tab:acdc}, where MCD and Dice were compared. Though both methods achieved comparable performance on mid-ventricle slices on UKBB data (Table \ref{motion_evaluation}), here the proposed method generalised better on data in unseen domains, with much higher registration accuracy (Table \ref{tab:acdc}). This is likely due to the benefit of the biomechanical regularisation, which enforces the generated deformations to be biomechanically plausible and less sensitive to the domain shift problem.

\begin{minipage}{.45\linewidth}
\centering
\includegraphics[width=.95\linewidth]{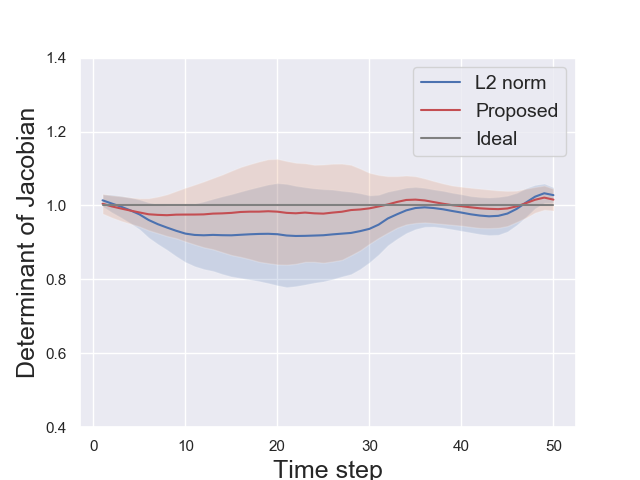}
\captionof{figure}{Determinant of Jacobian across cardiac cycle.}
\label{fig:jacobian}
\end{minipage}
\hfill
\begin{minipage}{.45\linewidth} 
    \centering
    \footnotesize
        \captionof{table}{Model generalisation performance on ACDC dataset.}
    \setlength{\tabcolsep}{2.5pt}
    \begin{tabular}{ccccc}
\toprule
\multirow{2}*{Method} &  \multicolumn{2}{c}{DL+L2}&\multicolumn{2}{c}{Proposed} \\
\cline{2-5}
 & {MCD}   & Dice  & MCD & Dice   \\
  \midrule
  {Apical} & 2.853  & 0.656 & \textbf{2.450}  & \textbf{0.707} \\
  \midrule
 {Mid} & 2.799  & 0.745 &\textbf{2.210}   & \textbf{0.783}\\
 \midrule
Basal & 2.814  & 0.751 & \textbf{2.229} & \textbf{0.789} \\
    \bottomrule
    \end{tabular}
    \label{tab:acdc}
\end{minipage}

\section{Conclusion}
In this paper, we have presented a novel biomechanics-informed neural network for image registration with an application to myocardial motion tracking in cardiac cine MRI. A VAE-based regulariser is proposed to exploit the biomechanical prior knowledge and to learn a manifold for biomechanically simulated deformations. Based on that, the biomechanics-informed network can be established by incorporating the learnt VAE regularisation to inform the registration process in generating plausible deformations. Experimental results have shown that the proposed method outperforms the other competing approaches that use conventional regularisation, achieving better motion tracking accuracy with more reasonable myocardial motion and strain estimates, as well as gaining better generalisability. For future work, we will extend the method for 3D motion tracking and investigate on incorporating physiology modelling.

\section*{Acknowledgements}
This work was supported by EPSRC programme grant SmartHeart (EP/P001009/1). This research has been conducted mainly using the UK Biobank Resource under Application Number 40119. The authors wish to thank all UK Biobank participants and staff.

\bibliographystyle{splncs04}
\bibliography{paper1103}

\newpage
\section*{Supplementary Material}
\beginsupplement
\vspace{-5mm}
\end{document}